\documentclass[12pt]{article}
\usepackage[dvips]{graphicx}
\begin{document}
\newcommand{\beq}{\begin{equation}}
\newcommand{\eeq}{\end{equation}}
\newcommand{\bdis}{\begin{displaymath}}
\newcommand{\edis}{\end{displaymath}}
\newcommand{\bea}{\begin{eqnarray}}
\newcommand{\eea}{\end{eqnarray}}
\newcommand{\barr}{\begin{array}}
\newcommand{\earr}{\end{array}}
\newcommand{\beas}{\begin{eqnarray*}}
\newcommand{\eeas}{\end{eqnarray*}}
\begin{center}
{\large{\bf New geometries associated with the nonlinear Schr\"{o}dinger equation}}
\vskip .4cm
{\bf S. Murugesh,}~~${\bf  Radha Balakrishnan}$
\vskip .3cm
{\it The Institute of Mathematical Sciences, Chennai 600 113, India}
\end{center}
\bigskip
\hrule
\bigskip
\baselineskip=20pt

\noindent {\bf Abstract}\\
 We apply our recent formalism
 establishing new connections
 between the geometry of moving space curves and soliton equations,
 to the nonlinear
 Schr\"{o}dinger equation (NLS).
  We show that  any given solution of the NLS gets associated with
  three  distinct space curve evolutions. The tangent
  vector of the first of these curves, the binormal vector of the second
  and the normal vector of the third, are shown to
  satisfy the integrable Landau-Lifshitz (LL) equation
  ${\bf S}_u = {\bf S} \times {\bf S}_{ss}$, (${\bf S}^2=1$).
 These  connections enable us to find the three surfaces
 swept out by the moving curves associated with
   the NLS.  As an example, surfaces  corresponding to  a
  stationary envelope  soliton solution of the NLS are obtained.

\noindent {\bf 1 Introduction}\\

\noindent A procedure to associate a completely
  integrable equation\cite{ablo}
 supporting soliton solutions
with the evolution equation of a moving space curve was found some
time ago by Lamb \cite{lamb}. Recently, we  showed \cite{muru} that
 there are two other distinct ways of making such a connection.
 Thus {\it three} different space curve evolutions
 get associated with
  a given solution of the integrable equation.
 As an illustrative example,
  we  considered the nonlinear Schr\"{o}dinger equation (NLS)
 and  demonstrated  that  the three associated moving curves
  had distinct  curvature  and torsion functions.
 We also obtained the curve  parameters for a one-soliton
 solution of the NLS.  However, as is well known\cite{stru},
 the  explicit construction of
 an  evolving space curve  or swept-out surface, using the
  corresponding expressions
 for the curvature and torsion  is a nontrivial task in general.
 For an  integrable nonlinear partial differential equation,
 a method proposed by Sym \cite{sym}
  shows  that using its Lax pair, a certain  surface that gets
 associated with a given  solution
  can be constructed, and this has been
 applied \cite{levi} to  the  NLS.
 In this paper,
 we use a different approach  which  obtains
 two more surfaces (or moving curves), in addition to the
 above surface. For the NLS, we first
 use  the   expressions\cite {muru} for the  associated
   curve parameters to
 show  that the three  space curve
 evolutions  {\it all}  map to the integrable Landau-Lifshitz (LL)
 equation\cite{land} for the
 time evolution of a  spin vector ${\bf S}$ of a continuous
 one-dimensional Heisenberg ferromagnet. In other words,
   the tangent vector of the first  moving curve, the binormal
 vector of  the second, and  the
  normal  vector of  the third,  are shown to satisfy the LL equation.
  The first of these results
  is  essentially the {\it converse}  of the
 important mapping  from the LL equation to the NLS
  which Lakshmanan\cite{laks} had obtained, by identifying {\bf S} with the
 tangent vector.
 Exploiting  the above  connections
   enables us to  explicitly construct the three
   swept-out surfaces. Surfaces associated with a stationary
 envelope soliton of the NLS  are presented.\\

\noindent {\bf 2 New connections between moving curves and soliton equations}\\

\noindent A moving space curve embedded in three dimensions
  may be described\cite{radha}
 using the following two sets of Frenet-Serret equations\cite{stru}
 for the orthonormal triad of unit vectors made up of the tangent
 ${\bf t}$, normal ${\bf n}$ and the binormal ${\bf b}$:
\beq\label{eq1}
{\bf t}_s \ = \ K {\bf n} \ ; \ {\bf n}_s  \ =
 \ -K{\bf t} \ + \ \tau
{\bf b}\ ; \ {\bf b}_s \ =\ -\tau {\bf n}.
\eeq
\beq\label{eq2}
{\bf{t}}_u\,= g {\bf {n}} +h  {\bf{b}}\ ;\   {\bf{n}}_u \,=
-  g {\bf{t}} + \tau_{0} {\bf{b}}\ ; \  {\bf{b}}_u\,=\,-h
{\bf{t}} - \tau_{0} {\bf{n}}.
\eeq
Here,  $s$ and  $u$ denote the arclength and time respectively.
The parameters  $K$ and $\tau$ represent the curvature and torsion
 of the space curve.
 The parameters $g,h$ and $\tau_0$ are, at this stage,
general parameters which determine
 the time evolution  of the curve. All the parameters are
  functions of both $s$ and $u$. The subscripts $s$ and $u$ stand for  partial derivatives.
On requiring the compatibility conditions
\beq
{\bf t}_{su} \ = \ {\bf t}_{us} \ ;
 \ {\bf n}_{su} \ = \ {\bf n}_{us}
\quad ; \quad {\bf b}_{su} \ = \ {\bf b}_{us},
\eeq
a short calculation using Eqs. (\ref{eq1}) and (\ref{eq2}) leads to
\beq\label{eq4}
K_u=(g_s-\tau h);~~\tau_u=(\tau_0)_s+Kh;~~h_s=(K\tau_0-\tau g).
\eeq
\noindent {\small {\bf Formulation I:}} We shall refer to Lamb's procedure
 \cite{lamb} for associating moving space curves with soliton equations
as "formulation I", to distinguish it from two others to follow.
 We remark that although  Eq.(\ref{eq2}) was  not
  introduced by Lamb, his formulation implied them. As we shall
 see, its explicit introduction \cite{radha} proves  very convenient in
 unraveling the geometry of the associated soliton equation.
  This formulation was motivated by Hasimoto's
 earlier work \cite{hasi}, which had established a connection
 between the local induction equation for a vortex
 filament in a fluid \cite{dari} and the NLS. Here, one proceeds by defining
  a  complex vector  ${\bf N}=({\bf n}+i{\bf b})~~ {\rm exp}[i\int \tau~ds]$
 and the Hasimoto function
\beq\label{eq6}
\psi(s,u)=K\exp{[i\int \tau~ds]}.
\eeq
  By writing ${\bf N}_s, {\bf t}_s, {\bf N}_u$ and ${\bf t}_u$
 in terms of ${\bf t}$ and ${\bf N}$, imposing the compatibility
 condition ${\bf N}_{su} \ = \ {\bf N}_{us}$,
and equating the coefficients of {\bf t} and {\bf N} in it,  one obtains
\beq\label{eq7}
\psi_u+\gamma_{1s}+(1/2)[\int~(\gamma_1\psi^*-\gamma_1^*\psi)~~ds]~\psi=0.
\eeq
where
\beq\label{eq10}
\gamma_{1}=-(g+ih)~~{\rm exp}[i\int\tau~ds]
\eeq
 The key step in Lamb's work is that
 an appropriate choice of $\gamma_1$ as a function of $\psi$
 and its derivatives can yield
  a known integrable equation for $\psi$. Comparing
 a solution of this equation with the Hasimoto function (\ref{eq6})
 yields the curvature $K$ and  torsion $\tau$ of the moving space
 curve. Next, using the above mentioned
 specific  choice of $\gamma_1$  in Eq. (\ref{eq10})  yields the curve
 evolution parameters  $g$ and $h$  as  some specific
 functions of $K$, $\tau$ and their derivatives. Knowing these,
 $\tau_0$ can also be found from the third equality in
  Eq. (\ref{eq4}). Thus a  set of parameters $K$, $\tau$,
 $g$, $h$ and $\tau_0$
 that  correspond to a given solution of the integrable
 equation has been found. In other words,  associated with this
 solution, there exists a  certain  moving space curve
 determined using Lamb's procedure.

 This raises the following question:  Is this the only
 possible  curve evolution that one can associate
 with an integrable equation,
 or are there others?
 We showed recently \cite{muru} that there are two other ways of making
  the association, which we call formulations II and III respectively,
  which lead to two other curve evolutions.\\
\noindent{\small {\bf Formulation (II):}}  Here, we
 combine   the first two
 equations in Eqs.(\ref{eq1}) to   show that
 a  complex vector
 ${\bf M}=({\bf n}-i{\bf t}){\rm exp}[i\int K~ds]$ and a complex function
\beq\label{eq11}
\Phi(s,u)=\tau\exp{[i\int K~ds]},
\eeq
 appear in a natural fashion.
  By writing ${\bf M}_s, {\bf b}_s, {\bf M}_u$ and ${\bf b}_u$
 in terms of ${\bf M}$ and ${\bf b}$,
 setting ${\bf M}_{su} \ = \ {\bf M}_{us}$, and equating the
 coefficients of {\bf b} and {\bf M}, respectively, we get
\beq\label{eq13}
\Phi_u+\gamma_{2s}+(1/2)[\int~(\gamma_2\Phi^*-\gamma_2^*\Phi)~~ds]~\Phi=0.
\eeq
where
\beq\label{eq15}
\gamma_{2}=-(\tau_{0}-ih)~~{\rm exp}[i\int K~ds],
\eeq
 The subscript $2$ is used on $\gamma$
 to indicate formulation (II).\\
\noindent {\small {\bf Formulation (III):}}
  Here,  we combine  the first and
 third equations of (\ref{eq1}),  leading to the appearance of a
   complex vector ${\bf P}=({\bf t}-i{\bf b})$, and a complex function $\chi$
 given by\cite{fn}
\beq\label{eq18}
\chi(s,u)=(K+i\tau).
\eeq
   Next, writing ${\bf P}_s, {\bf n}_s, {\bf P}_u$ and ${\bf n}_u$
 in terms of ${\bf P}$ and ${\bf n}$, imposing the compatibility
   condition  ${\bf P}_{su} \ = \
{\bf P}_{us}$, and equating the coefficients of {\bf n} and {\bf P},
respectively, we get
\beq\label{eq19}
\chi_u+\gamma_{3s}+(1/2)[\int~(\gamma_3\chi^*-\gamma_3^*\chi)~~ds]~\chi=0.
\eeq
where
\beq\label{eq21}
\gamma_{3}=-(g+i\tau_{0}).
\eeq
 Here, the subscript $3$
corresponds  to  formulation (III).    Since  Eq. (\ref{eq13}) and  Eq.
 (\ref{eq19}) have the same form as Lamb's  equation  (\ref{eq7}),
  it is clear that  for a suitable choice (see discussion following Eq. (\ref{eq10}))
 of $\gamma_2$  as a  function of $\Phi$ and its derivatives,
  and of $\gamma_3$  as a  function of $\chi$ and its derivatives,
   these equations  can become known integrable  equations
 for $\Phi$ and $\chi$ respectively.

 Collecting our results, we see
 from Eqs. (\ref{eq6}), (\ref{eq11}) and (\ref{eq18}) that   the
complex functions $\psi$, $\Phi$ and $\chi$ that satisfy the integrable equations
 in the three formulations  are different
functions of $K$ and $\tau$. Further,  we see from
 Eqs. (\ref{eq10}), (\ref{eq15}) and (\ref{eq21})
 that the
  complex quantities  $\gamma_1$, $\gamma_2$ and $\gamma_3$
  that arise in these formulations
  also  involve different combinations  of  the curve
 evolution parameters  $g,h$  and  $\tau_{0}$.
 Thus it is clear that these formulations indeed describe three
 {\it distinct}
classes of curve motion that
 can be associated with a given integrable
  equation. (Our analysis suggests that this
 association may  extend to some  partially
 integrable equations as well.) Next, we  apply these results to the
 NLS.\\

\noindent{\bf 3 Application to the NLS}\\

From our discussion given in the last section,
it is easy to verify that in the three formulations, the respective choices
\beq\label{eq23}
\gamma_1  = -i \psi_{s};~~\gamma_2= -i \Phi_s;~~\gamma_3= -i \chi_s,
\eeq
when used in Eqs. (\ref{eq7}), (\ref{eq13}) and (\ref{eq19}), lead to the NLS
\beq\label{eq24}
iq_u+q_{ss}+ \frac{1}{2}|q|^2 q  =0,
\eeq
  with $q$ identified with
  the complex functions $\psi, \Phi$ and $\chi$
 respectively. Now,  a general solution
 of Eq. (\ref{eq24}) is of the form $q=\rho \exp [i\theta]$.
 Equating this  with the  complex functions  defined
 in Eqs.(\ref{eq6}), (\ref{eq11}) and (\ref{eq18}) yields
  the   curvature and the torsion  of the  space curves
  that correspond to that solution of the NLS to be
${\bf (I)}~ \kappa_{1}=\rho,~~ \tau_{1}= {\theta}_s,~~
{\bf (II)}~ \kappa_{2}= {\theta}_s,~~\tau_{2}=\rho$ and $
{\bf (III)}~ \kappa_{3}=\rho \cos \theta,~~\tau_{3}= \rho \sin \theta$.
Thus clearly, three distinct space curves get associated
 with the NLS. However, even if  $K$ and $\tau$
 are known, to {\it solve} the Frenet-Serret equations (\ref{eq1}) to
 find the tangent {\bf t} of the  curve,
 (in order to construct from it,
 the corresponding position vector ${\bf r}(s,u)=\int {\bf t}~~ds$
 that  describes the  moving curve) is usually very
 cumbersome in general.
 In the present  context, we shall show
 that  a certain connection of the underlying curve evolutions
 of the NLS with
 the  integrable LL equation via three  distinct mappings
    enables  us to
 construct  these curves.

 To proceed, first we  equate
  the expressions for $\gamma_1$, $\gamma_2$ and
$\gamma_3$ given  in  Eq. (\ref{eq23}) with those given
 in Eqs. (\ref{eq10}), (\ref{eq15}) and (\ref{eq21}) and  obtain
 the following curve evolution parameters $g, h$ and $\tau_0$
 in the three  cases:\\ $ {\bf (I)}~~g_{1}=-\kappa_{1}\tau_{1};~~
h_{1}=\kappa_{1s};~~
\tau_{01}=
(\kappa_{1ss}/\kappa_{1})-\tau_{1}^{2},\\
{\bf (II)}~~ g_{2}=(\tau_{2ss}/\tau_{2}) -\kappa_{2}^2;~~
 h_{2}=-\tau_{2s};~~
\tau_{02}=-\kappa_{2}\tau_{2},\\
{\bf (III)}~~g_{3}=-\tau_{3s};~~~
h_{3}=(1/2)(\kappa_{3}^{2}+\tau_{3}^{2});~~
\tau_{03}=\kappa_{3s}$.\\
 Next,  substituting  these   expressions
 for each of the  formulations  appropriately
  in Eq. (\ref{eq2}), and using Eq. (\ref{eq1}), a short
 calculation \cite{muru} shows that
 the LL equation\cite{land}
\beq\label{eq28}
{\bf S}_u = {\bf S} \times {\bf S}_{ss};~~~ {\bf S}^{2}=1
\eeq
 is obtained in {\it every} case, i.e., for  the {\it tangent} ${\bf t}_{1}$
  of the moving space curve in the first formulation, for the
   {\it binormal} ${\bf b}_{2}$
  in the second, and by the  {\it normal}
  ${\bf n}_{3}$  in the third.
 Of the above, the first  is just the  converse  of Lakshmanan's
 mapping\cite{laks}  where, starting with the LL equation (\ref{eq28}),
  and identifying
 ${\bf S}$ with the tangent to a moving curve, it  becomes
 possible to  obtain the  NLS for $\psi$. The other two clearly
 represent new geometries connected with the NLS.
  Furthermore, the converses of  these  two
   also hold good, i.e., starting with (\ref{eq28})
 and identifying ${\bf S}$ with ${\bf b}$ and ${\bf n}$ successively,
   we can show that the NLS for $\Phi$ and $\chi$ are obtained, respectively.
 These are the two analogs of Lakshmanan's mapping.
    Next, we exploit  these  connections  with Eq. (\ref{eq28})
  to find the  moving space curves associated with the NLS.

 The LL equation (\ref{eq28}) has been shown
 to be completely integrable \cite{takh} and gauge equivalent\cite{zakh}
 to the NLS.  Its exact solutions  can be found \cite{takh,tjon}.
 We now show how  ${\bf r}_1$, ${\bf r}_2$ and ${\bf r}_3$, the
 position vectors generating the three moving curves
 underlying the NLS, can be found in terms of  an exact solution ${\bf S}$
 of Eq. (\ref{eq28}).

\noindent {\bf (I)} Let ${\bf t}_{1}$ be the tangent to a
 certain moving curve created by a position vector ${\bf r}_1(s,u)$.
Thus we set  ${\bf t}_1 = {\bf r}_{1s} ={\bf S}$, a solution of the LL equation.
 Now, the corresponding  triad $({\bf t}_1,{\bf n}_1,{\bf b}_1)$
 of this curve
  satisfies  the Frenet-Serret equations (\ref{eq1}) with
 curvature $\kappa_1$ and torsion $\tau_1$. In terms of
 ${\bf t}_1$ (and hence ${\bf S}$), these are given by the usual expressions
\beq\label{eq29}
\kappa_1=|{\bf t}_{1s}|=|{\bf S}_s|;~~\tau_1=\frac{{\bf t}_1.
({\bf t}_{1s}\times{\bf t}_{1ss})}{{\bf t}^2_{1s}}=\frac{{\bf S}.
 ({\bf S}_{s}\times{\bf S}_{ss})}{{\bf S}^2_{s}}
\eeq
Thus the underlying moving curve ${\bf r}_1(s,u)$
  in this formulation is simply given in terms of the solution ${\bf S}$ by
\beq\label{eq30}
{\bf r}_1(s,u) = \int{{\bf t}_1~ ds} = \int{{\bf S}(s,u)~ds}
\eeq
  The above expression for ${\bf r}_1$  is the surface that
 one obtains using Sym's\cite{sym} method.\\
\noindent {\bf (II)}  Let the
  binormal of some moving curve ${\bf r}_2(s,u)$ be denoted by ${\bf b}_2$.
  For this case, ${\bf b}_2 = {\bf S}$.
 Here, the tangent ${\bf t}_{2} =  {\bf r}_{2s}$.
The  triad $({\bf t}_2,{\bf n}_2,{\bf b}_2)$ satisfies Eq. (\ref{eq1}) with
  curvature $\kappa_2 = {\bf b}_2.({\bf b}_{2s}\times{\bf b}_{2ss})
/|{\bf b}_{2s}|^2 =\tau_1$ and torsion  $\tau_2=\kappa_1$. (See Eq.(\ref{eq29})).
  Using  ${\bf t}_2 = {\bf n}_2\times{\bf b}_2 =
-{\bf b}_2\times{\bf b}_{2s}/|{\bf b}_{2s}|
=-{\bf S}\times{\bf S}_{s}/|{\bf S}_{s}|$,
 the position vector ${\bf r}_{2}(s,u)$ generating
the  second moving curve  is found to be
\beq\label{eq31}
{\bf r}_2(s,u) = \int{{\bf t}_2~ds} = -\int{{\bf S}\times\frac{{\bf S}_s}{|{\bf S}_s|}~ds}
\eeq
\noindent {\bf (III)} Finally, let the normal of yet another
 moving curve ${\bf r}_{3}(s,u)$ be denoted by ${\bf n}_{3}$.
 So we have  ${\bf n}_{3}= {\bf S}$.
 The tangent of this curve
is ${\bf t}_{3}={\bf r}_{3s}$, and the triad
$({\bf t}_{3},{\bf n}_{3},{\bf b}_3)$ satisfies Eq. (\ref{eq1})
 with curvature $\kappa_3$ and torsion $\tau_3$. Here, clearly,
 we need the expressions for ${\bf t}_3$
 in terms of ${\bf n}_3$ and its derivatives.
  From Eq. (\ref{eq1})  for this case,
\beq\label{eq32}
(\kappa_3^2 + \tau_3^2){\bf t}_3 = \tau_3({\bf n}_3\times{\bf n}_{3s}) - \kappa_3{\bf n}_{3s}
\eeq
Next we find $\kappa_3$ and $\tau_3$ interms of ${\bf n}_3$  by  showing that
$ ({\bf n}_{3s})^2 = (\kappa_3^2 + \tau_3^2) = \kappa_{1}^2$
and
${\bf n}_3.({\bf n}_{3s}\times{\bf n}_{3ss})/|{\bf n}_{3s}|^2
 = \tau_1= \frac{d}{ds}(\tan^{-1}(\tau_3/\kappa_3))$.
 Using $|{\bf n}_{3s}| = \kappa_1$,  a short calculation yields
$\kappa_3 =\kappa_1 \cos \eta_1$  and
$\tau_3 =\kappa_1~~\sin \eta_1$, where
 $\eta_1=[\int \tau_1~~ds~~+~~c_{1}(u)]$.
Here, $c_{1}(u)$ is  a function of time $u$ , which
 can be  found  in terms of $\kappa_1$ and $\tau_1$ using
 the appropriate Eqs. (4)
 for $\kappa_{3u}$ and $\tau_{3u}$. These details will be given
 elsewhere. Substituting  the above values
 for $\kappa_3$ and $\tau_3$
 into Eq. (\ref{eq32}), and setting ${\bf n}_3 = {\bf S}$,
  the position vector ${\bf r}_{3}(s,u)$
  creating the third moving space curve  can be found
 to be
\beq\label{eq33}
{\bf r}_{3}(s,u)= \int {\bf t}_3~ds=\int\frac
{[({\bf S}\times{\bf S}_s)\sin\eta_1
 - {\bf S}_s~~\cos\eta_1])}{\kappa_{1}}~ds
\eeq

\noindent{\bf 4 Example: Soliton geometries}\\

Defining  three orthogonal unit vectors
${\bf\hat{e}}_1=\Big\{1,0,0\Big\},~{\bf\hat{e}}_2=
\Big\{0,\cos\eta,\sin\eta\Big\},~{\bf\hat{e}}_3=\Big
\{0,-\sin\eta,\cos\eta\Big\}$,
a soliton solution of the LL equation (\ref{eq28})  is given by
\bea\label{eq34}
\lefteqn{{\bf S}(s,u) = (1-\mu\nu {\rm sech}^2(\nu\xi)){\bf\hat{e}}_1 {} }\nonumber\\
& & {}+\mu\nu {\rm sech}(\nu\xi)\tanh(\nu\xi){\bf\hat{e}}_2-\mu\lambda{\rm sech}(\nu\xi){\bf\hat{e}}_3
\eea
where $\xi = (s-2\lambda u), \eta = (\lambda s + (\nu^2 - \lambda^2)u)$,
 and $\mu = 2\nu/(\nu^2 + \lambda^2)$. Here, $\nu$ and $\lambda$ are
arbitrary constants.  Using Eq.({\ref{eq34}) and our results
 of the previous section, the three moving curves
that correspond to the  soliton solution
$q=~\rho~\exp {i\theta}=~2\nu{\rm sech}(\nu\xi)~\exp{i\eta}$
 of the NLS (Eq. (\ref{eq24})) are found
 by substituting Eq. (\ref{eq34}) in Eqs. (\ref{eq30}),
  (\ref{eq31}) and
   (\ref{eq33}),  respectively.
 For the sake of illustration,
 let us consider the special case $\lambda=0$, which corresponds
  to the velocity of the  envelope of the NLS soliton being zero.
   We obtain  the  following three swept-out surfaces:
\bea\label{eq47}
{\bf (I)}~~{\bf r}_{1}=[s-(2/\nu) \rm tanh \nu s,
 (-2/\nu) \rm sech \nu s~ \cos \nu^{2}u, {}  \nonumber\\
 {} (-2/\nu)\rm sech \nu s~\sin \nu^{2} u]
\eea
 Note that $\kappa_1 =2 \nu {\rm sech}(\nu s)$
    and  $\tau_1 = 0.$ This surface is given in Fig. (1).
\beq\label{eq48}
{\bf (II)}~~{\bf r}_{2}= s~~~[~~ 0,~~~
 \sin \nu^{2}u,~~~ -\rm \cos \nu^{2} u~~]
\eeq
 Here, $\kappa_2 = 0$ and $ \tau_2=2\nu {\rm sech}(\nu s)$.
  For the sake of completeness,
we display this planar surface in Fig. (2).
\bea\label{eq49}
{\bf (III)}~~{\bf r}_{3}= [(2/\nu) {\rm sech}\nu s~\cos(\nu^2 u),~
 (s-(2/\nu)\tanh \nu s\nonumber\\
\cos^2(\nu^2 u)), - (2/\nu)\tanh \nu s\cos(\nu^2 u)\sin(\nu^2 u)]
\eea
 Here, $\kappa_3 = 2\nu{\rm sech}\nu s~~\cos \nu^{2}u$ and $\tau_3 =
 2\nu{\rm sech}\nu s~~\sin\nu^{2}u.$
  This surface is given in Fig. (3).

  For the case  $\lambda \ne 0$, the envelope of the NLS
 soliton  {\it moves}.
 Geometrically, this  motion can be shown to correspond to the
  "twisting out" of the surface
 in Fig. (1), around its  symmetry axis,
 and "stacking up"  of more such surfaces in a helical fashion along this axis.
  This leads to corresponding changes in
  Figs. (2) and (3) as well. The details of this will be presented
 elsewhere.

   Before we conclude, we  mention that the geometry underlying
 the  NLS  can also be studied
 by working with the
 {\it complex conjugates}
 of the complex vectors and functions that we used in the three
 formulations.
  These
 can be shown to lead to a
  mapping to the LL equation for ${\bf -t},{\bf -n}$ and ${\bf -b}$
 respectively.
 It can be verified that these merely yield  surfaces  which are created by
 the {\it negative} of the position vectors ${\bf r}_{i},  i=1,2,3$,
 which we  found in Section 3,  so that essentially
   no new surfaces result from these.
 Finally, while in the first formulation,
 it can be easily verified that
 the curve velocity ${\bf r}_{1u}$
   satisfies  the local induction equation\cite{dari}
 ${\bf r}_{1u}~=~\kappa_{1} {\bf b}_1$,
  the  velocities ${\bf r}_{2u}$ and ${\bf r}_{3u}$
  appearing in the other two formulations  can be shown to
 satisfy more complicated equations. These  general results
  on curve kinematics  and their ramifications are reported
  in \cite{muru1}.

\noindent{\bf Figure Captions:}
\begin{figure}[h]
\resizebox{0.5\textwidth}{!}{%
\includegraphics{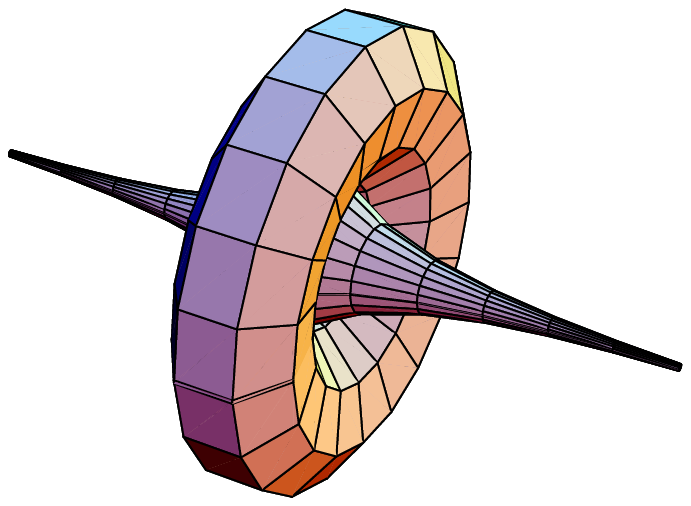}
}
\caption {Surface swept-out by the moving
 space curve ${\bf r}_{1}(s,u)$ (Eq. (\ref{eq47}))
 for $\nu=1$ and $0\le u\le 6.3$.}
 \end{figure}
\begin{figure}[h]
\resizebox{0.5\textwidth}{!}{%
\includegraphics{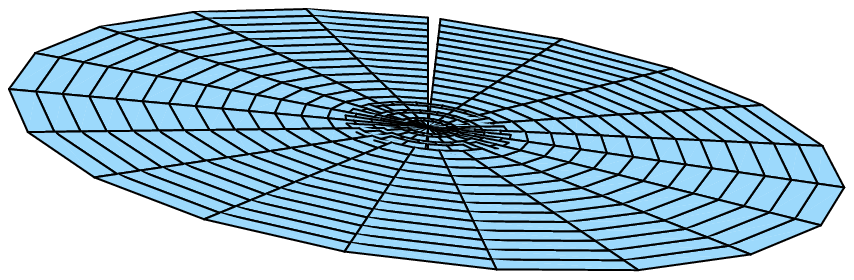}
}
\caption {Surface swept-out by the moving
 space curve ${\bf r}_{2}(s,u)$ (Eq. (\ref{eq48}))
 for $\nu=0.5$ and $0 \le u \le 25$.}
\end{figure}
\begin{figure}[t]
\resizebox{0.5\textwidth}{!}{%
\includegraphics{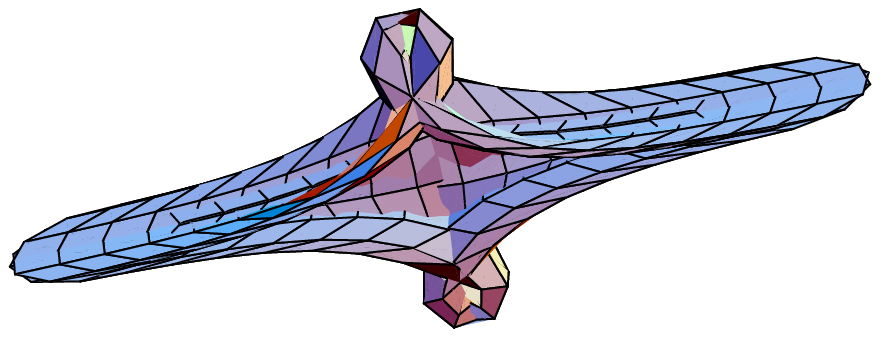}
}
\caption {Surface swept-out by the moving
 space curve ${\bf r}_{3}(s,u)$ (Eq. (\ref{eq49}))
 for $\nu=0.5$ and $0 \le u \le 25$.}
\end{figure}
\end{document}